\documentclass [preprint,aps,showpacs]{revtex4}
\usepackage[final]{graphics}
\usepackage{amssymb}
\usepackage{amsfonts}
\usepackage{epsfig}

\begin{document}
\draft
\title{Percolation on two- and three-dimensional lattices}
\author{P. H. L. Martins 
and J. A. Plascak\footnote{phlm@fisica.ufmg.br and pla@fisica.ufmg.br} }
\address{Departamento de F\'{\i}sica, Instituto de Ci\^encias Exatas,\\ 
Universidade Federal de Minas Gerais \\
C. P. 702, 30123-970, Belo Horizonte, MG - Brazil }
\date{\today}

\begin{abstract}

 In this work we apply a highly efficient Monte Carlo algorithm recently
proposed by Newman and Ziff to treat percolation problems. The site and
bond percolation are studied on a number of lattices in two and three 
dimensions. Quite good results for the wrapping
probabilities, correlation length critical exponent and critical
concentration are obtained for the square, simple cubic, HCP and 
hexagonal lattices by 
using relatively small systems. We also confirm the universal aspect
of the wrapping probabilities regarding site and bond dilution.

\end{abstract}

\pacs{02.70.-c, 05.10.Ln, 05.70.Jk, 64.60.Ak}

\maketitle

\section{Introduction}

Among the several methods for treating disordered systems and geometrical
problems,  percolation theory is certainly one of the most important. 
Due to its similarities with transitions that occur in many other systems 
(not only physical, but biological, social etc.), percolation has been used 
in studies within a large variety of fields. Forest fires \cite{Henley93}, 
biological evolution \cite{Ray94,Jovanovic94}, epidemics \cite{Moore00}, 
social influence \cite{Solomon00} and dilute magnetism \cite{Stinch} 
are only a few examples of the wide 
applicability of this theory, as well as percolation 
(e.g., in porous media) itself \cite{Stauffer91}.

Although easily  defined, percolation presents theoretical and computational
difficulties. For instance, the percolation
threshold for the site problem on a simple square lattice is not known exactly.
Therefore, approximate solutions are necessary, and much effort has 
been dedicated in this direction. From the theoretical point of view, one can 
utilize mean-field \cite{Gaveau93,Rudnick98} and  renormalization
group \cite{Young75,Burkhardt82,Sahimi95,pla} techniques, among others. 
In particular, computer simulations constitute a powerful tool in this area, 
since their application to percolation is simpler than for many other 
problems in statistical physics \cite{landau}. 
Typically, one can obtain a valid configuration by simply 
populating sites (or bonds) in a given lattice. To measure quantities of 
interest, such as the percolation threshold or the mean cluster size, it is 
necessary to identify all clusters in the configuration. For this purpose, 
many algorithms have been developed, the best known perhaps being that devised
by Hoshen and Kopelman \cite{Hoshen76}. Other algorithms, like 
hull-generation \cite{Ziff84,Grassberger92}, can also be used, but  
only to answer some specific questions. More recently, Newman and Ziff
proposed a new algorithm \cite{Newman00}, which is general and quite 
efficient, both in its computational requirements  and in its accuracy.
Although the algorithm can be used to obtain any  observable of the problem,
in their papers they have used it to calculate the so-called 
wrapping probabilities to investigate a number of aspects of the problem.
For example, using exact values of the wrapping 
probabilities, a high-precision result for the site percolation threshold 
on the square lattice was obtained \cite{Newman00}.

Unfortunately, 
for most lattices, one does not know the exact wrapping probabilities
and, in using them, the problem must be tackled in a different way.
One of the wrapping probabilities -
$R_L^{(1)}$, which is the probability that a cluster wraps around one specified
axis, but not around the other ones - is particularly useful. In this case 
it is not necessary to know its exact value at the critical concentration
for the infinite system since it has a maximum from which the critical point
can be obtained. The method, however, is indeed capable to properly estimate
the other wrapping probabilities which do not exhibit a maximum (in such
cases there is just a crossing region close to the critical threshold).
On the other hand, as 
we will see below, there are still some probabilities in dimensions higher 
than two that present, besides the maximum, a crossing region from where 
critical behavior is also achieved.

In this work we compute the percolation threshold and
the correlation length exponent, as well as the set of the unknown wrapping 
probabilities using the Newman-Ziff
algorithm. These quantities are evaluated by employing the usual 
finite-size scaling as well as a cell-to-cell scheme\cite{ziff02}. 
After summarizing the Newman-Ziff 
approach in the next section we describe, in section III, the
method which enables us to evaluate such geometrical quantities 
and we apply it to site and bond percolation on the square, simple cubic,
HCP (hexagonal close-packed) and simple hexagonal lattices. 
Concluding remarks are given in the final section.

\section{The Newman-Ziff algorithm}

To determine the percolation transition, this algorithm uses
the wrapping probability $R_L(p)$, which, for a given site (or bond) 
occupation $p$, is basically the probability that a
cluster wraps around a 
system with periodic boundary conditions on a lattice of linear dimension $L$. 
This wrapping can, however, be defined in various manners, each with its 
own probability. For instance, on two-dimensional lattices one has:
(i) $R_L^{(h)}$ and $R_L^{(v)}$, the probabilities that a cluster wraps
    around the system in the horizontal or vertical direction, respectively
    (on a square lattice these quantities are equal);
(ii) $R_L^{(e)}$, the probability that the cluster wraps around the lattice 
     $either$ horizontally $or$ vertically (or both);
(iii) $R_L^{(b)}$ for wrapping in $both$ horizontal $and$ vertical directions;
(iv) $R_L^{(1)}$ for wrapping around $one$ specified direction but $not$ 
     the other one.
Different lattices can allow for further geometrical choices for
$R_L(p)$. For example,
on a simple cubic lattice, besides $R_L^{(1)}$, we can define $R_L^{(2)}$ as
the probability that there exists a cluster that wraps the system in
two directions, but not around the third one. Analogously to $R_L^{(b)}$, 
we have $R_L^{(3)}$ for wrapping around the three directions. 

In order to evaluate these quantities it is necessary to generate many 
independent realizations
of the algorithm, each of them consisting of the following steps:\\
(1) Initially, all sites are empty;\\
(2) Sites are chosen to be occupied at random;\\
(3) When a new site is added, one must check all its neighbors to 
verify if
the new site forms an isolated cluster (all neighbors empty) or if it joins
together two or more clusters. In the first case, we need do nothing. 
In the latter, we have to update the cluster listing.
Clusters are stored in a tree structure, with one site of each cluster  
considered  the $root$ site. All sites in a given cluster, other than the 
root, have a pointer to some other site in the same cluster, such that by
following a succession of such pointers one can ultimately reach the root. 
In order to join two clusters we simply add a pointer 
from the root of the smaller cluster to the root of the larger one;\\
(4) Each time step (3) is repeated, we evaluate the
quantities of interest $Q_L^n$ as a function of the number $n$ of 
occupied sites.
$Q_L^n$ may be any of the wrapping probabilities $R_L$. 
Let $n^{\prime}$ be the
number of occupied sites for which percolation first occurs in a given 
realization.
$Q_L^n$ represents the fraction 
of realizations in which $n^{\prime}$ is less than or equal to $n$. Using all 
$Q_L^n$'s so evaluated, it is possible to calculate the function $Q_L(p)$ for
all possible values of $p$ in the range between $0$ and $1$ by a 
convolution  with the binomial distribution \cite{Newman00}:
\begin{equation}
Q_L(p)={\sum_n {N\choose n} p^n(1-p)^{N-n}Q_L^n}~.\label{Qp}
\end{equation}
For the bond percolation problem, we just replace sites by bonds in the above
steps.
 
The evaluation of the statistical errors can be done in a conventional
fashion. As discussed  in reference \cite{Newman00} the standard
deviation of the binomial distribution (\ref{Qp}) is given by
\begin{equation}
\sigma_{Q_L(p)}=\sqrt{{Q_L(p)\left[1-Q_L(p)\right]}\over {MCS}}~,\label{erro}
\end{equation}
where $Q_L(p)$ in the above equation has been taken as the mean value of the
corresponding wrapping probability and $MCS$ is the number of
Monte Carlo steps per site.

\begin{figure}[ht]
\includegraphics[clip,angle=0,width=7.9cm]{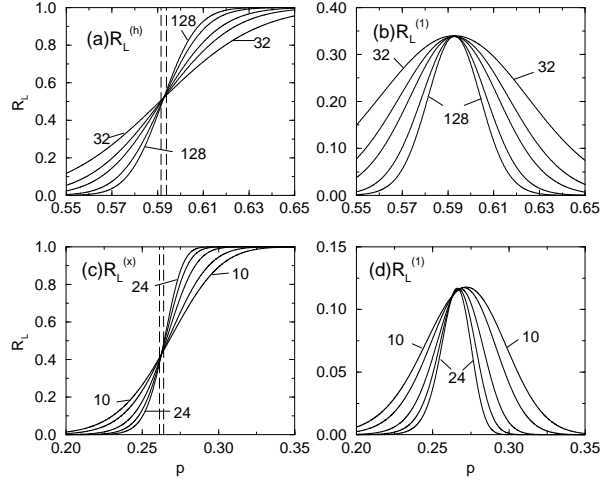}
\caption{\label{Fig1}Wrapping probabilities $R_L$ as a
                     function of occupation probability for site percolation 
                     on: square lattices
(a) $R_L^{(h)}$ and (b) $R_L^{(1)}$; and 
hexagonal lattices (c) $R_L^{(x)}$ and (d) $R_L^{(1)}$.   
For the square lattice the data have been obtained by taking
$4.0 \times 10^5$ MCS and lattice sizes {\it L} = 32, 48, 64, 96, 128.
For the hexagonal lattice the data have been obtained by taking
$5.0 \times 10^5$ MCS and lattice sizes {\it L} = 10, 12, 16, 20, 24.
In all figures the error bars were omitted for a better visualization.
Vertical dashed lines represent the lowest 
and the greatest  values of $p$ we have 
utilized for evaluating wrapping probabilities.
}
\end{figure}

As an illustration, we show in Fig. \ref{Fig1}(a) and \ref{Fig1}(b) 
the wrapping probabilities 
$R_L^{(h)}$ and $R_L^{(1)}$ as a function of the concentration $p$ of occupied 
sites for square lattices of different sizes. 
The exact values $R_{\infty}(p_c)$ of these quantities for an infinite 
square system were derived by Pinson \cite{Pinson94} and Ziff \cite{Ziff99}. 
In Ref. \cite{Newman00} it has been used these exact values to obtain an estimate 
for the percolation threshold $p_c$. For each system size $L$,
one determines the $p$ value yielding a wrapping probability  
equal to exact critical value. This $p$ value provides the estimate 
of $p_c^L$ for that $L$. In the critical region, one knows that
the estimates $p_c^L$ converge to the threshold $p_c$ according to
%

%
\begin{equation}
p_c^L - p_c \sim L^{-\theta-1/\nu} \label{eq1}.
\end{equation}
For square systems, using the known value $4/3$ of the exponent $\nu$ and
$\theta=2$, as obtained in \cite{Newman00}, we have
$p_c^L - p_c \sim L^{-11/4}$. By a finite size scaling, Newman and Ziff 
obtained $p_c=0.592~746~21(13)$
for the infinite system. This procedure is more complicated
in higher dimensions. Since neither $\theta$ nor $\nu$ are known, one has to
vary the scaling exponent to obtain a straight line. In Ref. \cite{Newman00},
it was found that the estimates of the percolation threshold for a simple
cubic lattice scale as $L^{-2}$. Thus, it seems to be not so easy to determine $\nu$
and $\theta$ separately by this method.

One has to note (Fig. \ref{Fig1}b) that $R_L^{(1)}$ is  different from
the other probabilities, as it exhibits a maximum. In this case, $p_c^L$ can
be estimated from the position of this maximum. $R_L^{(1)}$ is then of  
particular utility in systems for which the exact values are not known. 
We will see, moreover, that all the other wrapping probabilities 
can also be used to estimate the percolation threshold, as well as 
the correlation length critical exponent, on  any lattice. There are,
in addition, some quantities like $R_L^{(2)}$ in three dimensions which
exhibit both a maximum and a crossing region.

\section{approach and results}

We have applied the Newman-Ziff algorithm to site percolation on the
two-dimensional ($2d$)  square, the three-dimensional simple hexagonal
and simple cubic  ($3d$) lattices, and to bond percolation on square (2$d$), 
simple cubic  and HCP (hexagonal close-packed) lattices. 
Table \ref{Tab1} gives the system and sample sizes used
in our study.

\begin{figure}[ht]
\includegraphics[clip,angle=0,width=7.3cm]{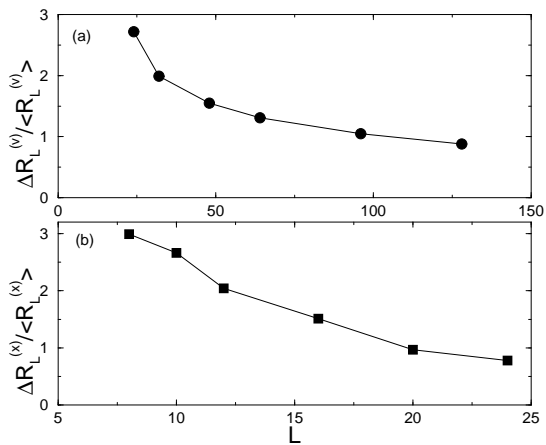}
\caption{\label{Fig1.5}(a) Relative error of $R_L^{(v)}$, where 
$\Delta R_L^{(v)}=\sigma_{R_L^{v}} $ 
for site percolation on the square lattice. 
(b) Relative error of $R_L^{(x)}$, where 
$\Delta R_L^{(x)}=\sigma_{R_L^{x}}$ 
for site percolation on the hexagonal lattice.
}
\end{figure}

Before discussing the results, we analyze the standard deviation of some 
particular quantities. Fig. \ref{Fig1.5} shows the relative error of $R_L^{(v)}$ 
for the square lattice and $R_L^{(x)}$ for the hexagonal lattice as a 
function of the $L$, at the critical concentration. 
Apart from a strong dependence with small $L$,
we clearly see that for larger lattice sizes the
relative error is almost independent of $L$, as predicted by Eq. (\ref{erro})
and previously stressed by
Newman and Ziff \cite{Newman00}, even for the three-dimensional 
hexagonal lattice. Similar behavior is found for other wrapping
probabilities, other three-dimensional lattices, as well as for the bond 
problem in different lattice structures. 

\begin{figure}[ht]
\includegraphics[clip,angle=0,width=7.3cm]{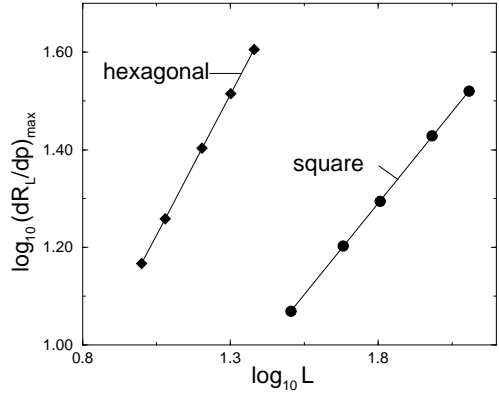}
\caption{\label{Fig2}Maximum derivative of 
wrapping probabilities ($R_L^{(v)}$ for  site percolation on the square 
lattice and $R_L^{(x)}$ for site percolation on the hexagonal 
lattice) as a function of  system size. 
Error bars are smaller than the symbol 
sizes. Linear regression of the data gives 
$\nu=1.334(4)$ for the square lattice  and $\nu=0.866(1)$ for the 
hexagonal lattice.
}
\end{figure}

Let us now discuss the evaluation of the critical exponent, the percolation
threshold and the wrapping probabilities. In order to get an idea
of the performance of the present approach, we will first apply it to the
problem in two dimensions where exact (or more accurate) results are
available.
From Fig. \ref{Fig1}(a) one sees that the derivative of $R_L^{(h)}$ at
the critical concentration increases as the lattice size increases.
In fact, one expects
that the maximum derivative of any wrapping probability not exhibiting 
a maximum scales as \cite{hu} 
\begin{equation}
\Big( \frac{{\rm d}R_L}{{\rm d}p} \Big)_{max} \sim L^{1/\nu}.
\label{derscale}
\end{equation}
Thus, the critical exponent $\nu$ can be estimated without any consideration
of the critical concentration $p_c$ by taking the scaling behavior of the
derivatives of the thermodynamic quantities $R_L$. They can  be 
straightforwardly computed from relation (\ref{Qp})
\begin{eqnarray}
\frac{{\rm d}Q_L}{{\rm d}p} = \sum_n \Bigg[ {N \choose n} n p^{n-1} 
(1-p)^{N-n} \nonumber
\end{eqnarray}
\begin{eqnarray}
 - {N \choose n} (N-n) p^{n} (1-p)^{N-n-1} \Bigg] Q_L^n~.
\end{eqnarray} 
In Fig. \ref{Fig2} we plot, on $\log _{10}$ scales, the maximum value of the 
derivative of $R_L^{(v)}$ as a function of system size,
for site percolation on the square lattice  
(the hexagonal lattice will be discussed later); 
a linear fit yields $\nu^{(v)}=1.334(4)$. Other quantities give
independent estimates of the exponent
(the corresponding data are too close to those of
$R_L^{(v)}$ to be distinguished on the scale of Fig. \ref{Fig2}).  
We find  $\nu^{(h)}=1.331(2)$; 
 $\nu^{(b)}=1.339(4)$ and $\nu^{(e)}=1.327(1)$.
Combining these four estimates we obtain  $\nu=1.333(5)$, in very good
agreement with the exact result $4/3$.

\begin{figure}[ht]
\includegraphics[clip,angle=0,width=7.3cm]{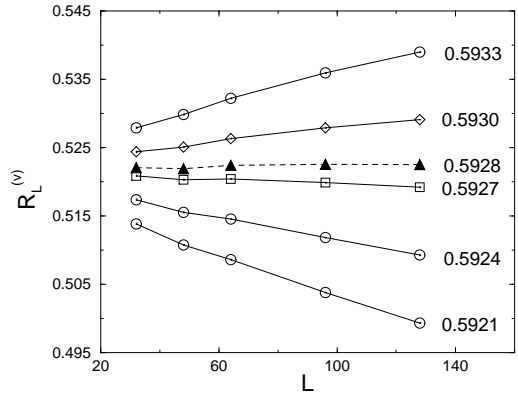}
\caption{\label{Fig3}Wrapping probability $R_L^{(v)}$ as a function of
the lattice size $L$ for site percolation on square lattices. 
Different lines correspond to  different concentrations {\it p}. 
 The best constant horizontal
line is represented by full triangles 
yielding {$ p_c$} = 0.5928(2) and
$R_L^{(v)}=0.523(4)$. Error bars are smaller than the symbol sizes. 
}
\end{figure}

Fig. \ref{Fig3} illustrates the approach for evaluating the critical
concentration, as well as the wrapping probabilities at $p_c$, applied again
to the site percolation problem on the square lattice, through a cell-to-cell
estimate. 
For a fixed probability occupation $p$, we compute the specified wrapping 
probability as a function of the lattice size. Fig. \ref{Fig3}  shows
$R_L^{(v)}$ as a function  of $L$.
For $p<p_c$,   $R_L^{(v)}$
decreases with increasing lattice size. 
For $p>p_c$, it increases. 
%
%
Exactly at $p_c$ one expects the  wrapping probability
to be independent of system size. Thus, by varying  $p$ in the critical
region and  searching for a constant $R_L^{(v)}$ we obtain an estimate 
for $p_c$. The vertical dashed lines in Fig. \ref{Fig1}(a) represent the 
limits on the $p$ values studied. From the data for 
$R_L^{(v)}$ in Fig. \ref{Fig3}, we have then
$p_c = 0.5928(2)$ and $R_L^{(v)}=0.523(4)$. 
Combining this estimate with the ones coming from the
other quantities we obtain the values listed in Table \ref{Tab2}. 
The results are quite close to the exact or expected ones, despite the 
small systems (sizes up to $128 \times 128$) and short runs 
(see table \ref{Tab1}).  Table \ref{Tab2} also gives the results obtained from
the present procedure to the bond percolation problem (which is easily
implemented in the algorithm) with an excellent estimate
of the known critical concentration. Moreover, the wrapping 
probabilities, within the error bars, are the same for site 
and bond problems, confirming the universal character of these
quantities.

Having demonstrated the  good performance of the method in cases
where exact results are available, we  studied some three-dimensional
lattices where data are not so ubiquitous in the literature. In particular,
we treat the simple cubic, simple hexagonal and the HCP (hexagonal 
close-packed) lattices.
To our knowledge, there are no results available  for 
the wrapping probabilities on such geometries as well as no indication
of their universal aspect regarding site and bond dilution.

\begin{figure}[ht]
\includegraphics[clip,angle=0,width=7.3cm]{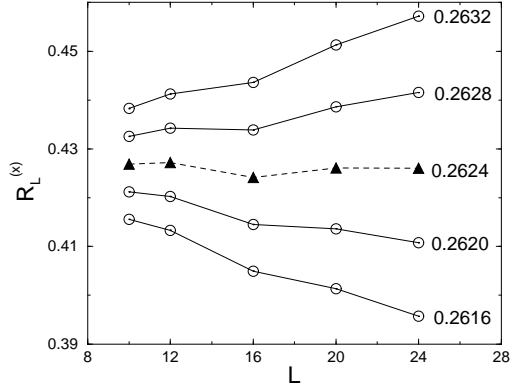}
\caption{\label{Fig4}Wrapping probability $R_L^{(x)}$ as a function of
the lattice size $L$ for site percolation on hexagonal lattices. 
Different lines correspond to different concentrations {\it p}. 
The best horizontal
line is represented by full triangles, yielding 
{$ p_c$} = 0.2624(4) and
$R_L^{(x)}=0.429(5)$. Error bars are smaller than the symbol sizes.
}
\end{figure}

As an example, we show in Fig. \ref{Fig1}(c) and \ref{Fig1}(d) 
the wrapping probabilities 
$R_L^{(x)}$ and $R_L^{(1)}$ as a function of $p$ for various lattice 
sizes, for site percolation on the simple hexagonal lattice. 
The corresponding scaling behavior of the derivative of 
$R_L^{(x)}$ is depicted in Fig. \ref{Fig2}; an estimate for 
the critical exponent $\nu$ may be extracted from these data. 
In Fig. \ref{Fig4} we show the 
estimate for $p_c$ as
well as the value of the wrapping probability $R_L^{(x)}$. The combined
results are listed in Table \ref{Tab3} together with the values for the
simple cubic lattice and those
obtained for the bond percolation on the HCP and simple cubic lattices. 
One can clearly see
that the wrapping probabilities are distinct for different
geometries, as is the critical concentration. Not only are our
$p_c$ estimate comparable to the values obtained previously, but
the critical exponents found here are close to the expected result for this
universality class, namely  $\nu=0.83(5)$ from series \cite{Dunn},
$\nu=0.88(1)$\cite{Stauffer91}, $\nu=0.8765(16)$\cite{ball} and
$\nu=0.893(40)$\cite{tomita} from Monte Carlo simulations.
%

The data of Table \ref{Tab3} are, up to our knowledge, quite new for these
three dimensional lattices. A by-product of the present results concerns
the universality of $R_L(p_c)$ at the percolation threshold. One can clearly
see that for site or bond percolation the wrapping probabilities of the
simple cubic lattice are, within the error bars, the same (as well as for the
problem in two dimensions depicted in Table \ref{Tab2}). This is in quite
good agreement with the expected universal aspect previously obtained for the 
spanning probability in general
dimensions and with both free and periodic boundary conditions \cite{hovi}.

\begin{figure}[ht]
\includegraphics[clip,angle=0,width=7.3cm]{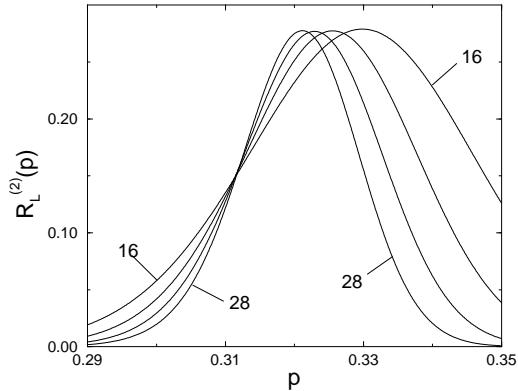}
\caption{\label{Fig6}Wrapping probability $R_L^{(2)}(p)$ as a function of $p$ 
for the site percolation on the simple cubic lattice. From the position of the 
peaks one estimates $p_c=0.3171(3) $ and from the procedure of the crossing 
region one gets $p_c=0.3116(3)$.
}
\end{figure}

Another interesting aspect of the three-dimensional lattices is the 
behavior of the quantity  $R_L^{(2)}(p)$ giving the probability of wrapping
around two directions and not around the third direction. Fig. \ref{Fig6}
shows such behavior for the simple cubic site diluted problem. In this case
one can obtain an estimate of the critical concentration not only from the 
position of its 
maximum but also from the crossings at $p_c$. It is noted, however,  that
a more accurate value is achieved from the analysis of the crossings (as done
in Fig. \ref{Fig4}) than from the location of its maximum. The same behavior
occurs for the other lattice structures.

%
%

\section{Concluding remarks}

We have seen that the results obtained using the present method are in good 
agreement with the exact (when available) or expected ones.
It is important to note that this procedure has been implemented using
relatively small systems and short Monte Carlo runs; better results could be 
achieved in larger-scale simulations. 
In addition, from the computed wrapping probabilities
at $p_c$ for the square and simple cubic lattices one can also confirm their 
universal  aspect regarding site and bond dilution. The same should hold, 
of course, for the other lattice geometries.

A slightly modified version of the present procedure, not using the cell-to-cell
estimate, can also be
applied to problems on two-dimensional lattices. Instead of tuning $p$, 
one can tune one of the $R_L$ in the critical 
region. For a given quantity $R_L$ (say, for example $R_L^{(h)}$) we fix it 
at a specified value $R^*$ on the vicinity of the critical point, and 
proceed analogously to Newman and Ziff's original approach. 
Observe that $R^*$ is in this case a
first estimate for $R_{\infty}(p_c)$. One can then compute, for each $L$, 
the intercept
between the function $R_L(p)$, previously evaluated, with the fixed $R^*$. 
Each of these intercepts gives an estimate for $p_c^L$, which is expected, 
for two-dimensional
systems, to scale  as $L^{-\phi}$ with $\phi=11/4$. Therefore, plotting 
$p_c^L \times L^{-11/4}$ for
different values of $R^*$, we can estimate the true $R_{\infty}$ as well as 
the percolation threshold by looking for the value of $R^*$ that yields the 
best straight line. Table \ref{Tab2} reports the critical values
so obtained for the two-dimensional lattice. The results are quite
good and comparable to those obtained in the previous section for both
the wrapping probabilities and the critical concentration. However,
the errors are in general considerably larger. 
In three dimensions this procedure can be implemented only if we know
the exponent $\phi$ beforehand.

\acknowledgments
{
Fruitful discussions with R. Dickman and D. P. Landau are gratefully
acknowledged. We would also like to thank the referee for several comments
on this work. 
This research was supported in part by Conselho Nacional de Desenvolvimento
Cient\'{\i}fico e Tecnol\'ogico (CNPq), Funda\c c\~ao Coordena\c c\~ao de 
Aperfei\c{c}oamento de Pessoal de N\'{\i}vel Superior (CAPES) and
Funda\c c\~ao de Amparo \`a Pesquisa do Estado de Minas Gerais (FAPEMIG)
 - Brazilian Agencies.
}

\newpage

\newpage

\begin{table}
\caption{Lattice sizes and run lengths (MCS) used in this work.
         The smallest and largest figures correspond to the total number
         of sites or bonds. In parenthesis we have the corresponding 
         lattice size $L$. } 
\begin{tabular}{cccc}
\tableline
\empty{Lattice} & \empty{Smallest} & \empty{Largest} & 
\empty{MCS ($10^5$)}  \\ 
\tableline
\multicolumn{4}{c}{Site percolation} \\
\tableline
Square     & 1 024(32) &  16 384(128)  &  4.0 \\     
Hexagonal  &  4 096(16)&  21 952(28)  & 1.0--10.0 \\    
Cubic      &  4 096(16)&  21 952(28)  & 1.0--2.0 \\
\tableline
\multicolumn{4}{c}{Bond Percolation} \\
\tableline
Square  & 2 048(32) &  32 768(128)  & 2.0--4.0 \\
HCP   &  768(4)  &  20 736(12)  & 5.0 \\
Cubic      & 12 288(16) &  52 728(26)  & 0.1--0.5 \\
\tableline 
\end{tabular}

\label{Tab1}
\end{table}\begin{table}
\caption{Results for site and bond percolation on the square lattice. 
Errors in parenthesis affect the last digits. For each case,
the first row shows the results
described in section III and the second row those from the modified approach 
briefly discussed in section IV (in the latter method 
the exact value for the exponent $\nu$) is used.}
\begin{tabular}{cccccccc}
\tableline
\multicolumn{8}{c}{Two dimensions} \\ \tableline
\empty{Lattice} & \empty{$R^{(h)}$} & \empty{$R^{(v)}$} & \empty{$R^{(e)}$} & 
\empty{$R^{(b)}$} & \empty{$\nu$} & \empty{$p_c$}(this work) & \empty{$p_c$} \\ \tableline
\empty{site} & 0.517(4) & 0.523(4) & 0.692(3) & 0.347(4) & 1.333(5) & 
0.592~7(1)  & 0.592~7\tablenotemark[1] \\ 

 & 0.521(9) & 0.524(6) & 0.695(7) & 0.353(5) & - & 
0.592~9(3)  & 0.592~7\tablenotemark[1] \\ 

\empty{bond} & 0.521(2) & 0.518(3) & 0.691(3) & 0.351(2) & 1.331(3)  & 
0.499~95(15) & 1/2 \tablenotemark[2]  \\ 

 & 0.517(11) & 0.519(13) & 0.684(16) & 0.348(6) & -  & 
0.499~8(4) & 1/2 \tablenotemark[2]  \\

\empty{Exact\tablenotemark[1]} & 0.5211 & 0.5211& 0.6905    & 0.3516& 
4/3 &  -  &   -  \\
\tableline 
\end{tabular}

\tablenotemark[1]{Ref. \cite{Newman00}. }\\
\tablenotemark[2]{Ref. \cite{Sykes}. } 
\label{Tab2}
\end{table}

\begin{table}
\caption{Results for site(s) percolation on hexagonal and simple
cubic lattices and for bond(b) percolation on simple cubic  and HCP lattices.
Errors in parenthesis affect the last digits.}
\begin{tabular}{ccccccccc}
\tableline 
\multicolumn{9}{c}{Three dimensions} \\ \tableline
\empty{Lattice} & \empty{$R^{(x)}$} & \empty{$R^{(y)}$} & \empty{$R^{(z)}$} & \empty{$R^{(e)}$} & 
\empty{$R^{(3)}$} & \empty{$\nu$} & \empty{$p_c$(this work)} & \empty{$p_c$} \\ \tableline
Hexagonal(s)& 0.429(5) & 0.332(5) & 0.183(4) & 0.467(6) & 0.120(3) & 0.867(14) & 
0.262~5(2) & 0.262~3(2)\tablenotemark[1] \\ 
Simple cubic(s)& 0.254(5) & 0.255(5) & 0.254(5) & 0.456(7) & 0.078(3) & 0.877(12) & 
0.311~5(3) & 0.311~6063(9)\tablenotemark[2] \\
Simple cubic(b)& 0.265(6) & 0.266(6) & 0.265(6) & 0.471(8) & 0.084(4) & 0.868(11) & 
0.249~0(2) & 0.248~8126(5)\tablenotemark[3] \\
HCP(b)     & 0.331(6) & 0.443(6) & 0.093(3) & 0.561(7) & 0.052(3) & 0.848(33) & 
0.120~3(2) & 0.119~9(2)\tablenotemark[1]  \\
\tableline 
\end{tabular}

\tablenotemark[1]{Ref. \cite{Marck97}.}\\
\tablenotemark[2]{Ref. \cite{ball}.}\\
\tablenotemark[3]{Ref. \cite{lorenz}.}
\label{Tab3}
\end{table}


\begin{thebibliography}{99}

\bibitem{Henley93}
C.~L.~Henley, Phys.~Rev.~Lett. {\bf 71}, 2741 (1993).

\bibitem{Ray94}
T. S. Ray and N. Jan, Phys.~Rev.~Lett. {\bf 72}, 4045 (1994).

\bibitem{Jovanovic94}
B.~Jovanovic, S.~V.~Buldyrev, S.~Havlin and H.~E.~Stanley, Phys.~Rev.~E
{\bf 50}, 2403 (1994).

\bibitem{Moore00}
C.~Moore and M.~E.~J.~Newman, Phys.~Rev.~E {\bf 61}, 5678 (2000).

\bibitem{Solomon00}
S.~Solomon, G.~Weisbuch, L.~deArcangelis, N.~Jan and D.~Stauffer, Physica~A
{\bf 277}, 239 (2000).

\bibitem{Stinch}
R. Stinchcombe in {\it Phase Transitions and Critical Phenomena} vol. 7,
eds. C. Domb and J. L. Lebowitz (London, Academic, 1983). See also D. P.
Belanger, Braz. J. Phys. {\bf 30}, 682 (2000).

\bibitem{Stauffer91}
D.~Stauffer and A.~Aharony, {\it Introduction to Percolation Theory}
(Taylor \& Francis, 1991).

\bibitem{Gaveau93}
B.~Gaveau and L.~S.~Schulman, J.~Stat.~Phys. {\bf 70}, 613 (1993).

\bibitem{Rudnick98}
J.~Rudnick, P.~Nakmahachalasint and G.~Gaspari, Phys.~Rev.~E {\bf 58},
5596 (1998).

\bibitem{Young75}
A.~P.~Young and R.~B.~Stinchcombe, J.~Phys. C {\bf 8}, L535 (1975).

\bibitem{Burkhardt82}
T.~W.~Burkhardt and J.~M.~J.~van Leeuwen, {\it Real Space Renormalization}
(Springer-Verlag, 1982).

\bibitem{Sahimi95}
M.~Sahimi and H.~Rassamdana, J.~Stat.~Phys {\bf 78}, 1157 (1995).

\bibitem{pla} J. A. Plascak, W. Figueiredo and B. C. S. Grandi,
Braz. J. Phys. {\bf 29}, 579 (1999).

\bibitem{landau} D. P. Landau and K. Binder, {\it A Guide to Monte Carlo 
Simulation in Statistical Physics}, Cambridge University Press (Cambridge, 
2000). 

\bibitem{Hoshen76}
J.~Hoshen and R.~Kopelman, Phys.~Rev.~B {\bf 14}, 3438 (1976).

\bibitem{Ziff84}
R.~M.~Ziff, P.~T.~Cummings and G.~Stell, J.~Phys. A {\bf 17}, 3009 (1984).

\bibitem{Grassberger92}
P.~Grassberger, J.~Phys. A {\bf 25}, 5475 (1992).

\bibitem{Newman00}
M.~E.~J.~Newman and R.~M.~Ziff, Phys.~Rev.~Lett. {\bf 85}, 4104 (2000);
 Phys.~Rev.~E {\bf 64}, 016706 (2001).

\bibitem{ziff02} see, for instance, R.~M.~Ziff and M.~E.~J.~Newman, 
 Phys. Rev E {\bf 66},  016129 (2002).

\bibitem{Pinson94}
H.~T.~Pinson, J.~Stat.~Phys. {\bf 75}, 1167 (1994).

\bibitem{Ziff99}
R.~M.~Ziff, C.~D.~Lorenz and P.~Kleban, Physica A {\bf 266}, 17 (1999).

\bibitem{hu}
C.-K. Hu, C.-Y. Lin and J.-A Chen, Phys.~Rev.~Lett. {\bf 75}, 193 (1995);
{\bf 75}, 2786(E) (1995).

\bibitem{Dunn} A. G. Dunn, J. W. Essam and D. S. Ritchie, J. Phys. C {\bf 8},
4219 (1975).

\bibitem{Sykes} M. F. Sykes and J. W. Essam, Phys. Rev. Lett. {\bf 10}, 3 (1963).

\bibitem{Marck97}
S.~C.~van~der~Marck, Phys.~Rev.~E {\bf 55}, 1514 (1997).

\bibitem{ball} H. G. Ballesteros, L. A. Fern\'andez, V. Mart\'{\i}n-Mayor,
A. Mu\~noz Sudupe, G. Parisi and J. J. Ruiz-Lorenzo, J. Phys. A: Math. Gen.
{\bf 32}, 1 (1999).

\bibitem{tomita} Y. Tomita and Y. Okabe, J. Phys. Soc. Jpn. {\bf 71 }, 1570
 (2002).

\bibitem{lorenz} C. D. Lorenz and R. M. Ziff, Phys. Rev. E {\bf 57}, 230 (1998).

\bibitem{hovi} J.-P. Hovi and A. Aharony, Phys. Rev. E {\bf 53}, 235 (1996).
\end{thebibliography}
\end{document}